# Analysis of Driver's Head Tilt Using a Mathematical Model of Motion Sickness

Takahiro Wada[a], Satoru Fujisawa[b], Shunichi Doi[b]

[a] College of Information Science and Engineering, Ritsumeikan University,
  1-1-1 Noji-higasi, Kusatsu, Shiga 525-8577, Japan
  twada@fc.ritsumei.ac.jp
[b] Faculty of Engineering, Kagawa University,
  2217-20 Hayashi-cho, Takamatsu, Kagawa 761-0396, Japan

Corresponding author:
  Takahiro Wada
  Tel & Fax:+81-77-561-2798
  Email: twada@fc.ritsumei.ac.jp
  College of Information Science and Engineering, Ritsumeikan University,
  1-1-1 Noji-higasi, Kusatsu, Shiga 525-8577, Japan

Abstract

  It is known that car drivers tilt their head toward the center of a curve. In addition, drivers are generally less susceptible to carsickness than are the passengers. This paper uses a mathematical model to investigate the effect of the head-tilt strategy on motion sickness. It is shown that tilting the head in the centripetal direction reduces the estimated motion sickness incidence (MSI), defined as the percentage of subjects who vomited. In addition, the head movements of both drivers and passengers were measured in a real car. It is also shown that the estimated MSI of the drivers is smaller than that of the passengers.

Experimental results presented in previous studies demonstrated that the severity of motion sickness was reduced when passengers imitated the head tilt of the driver. These results strongly suggest that the driver's head tilt reduces motion sickness, and this can be understood as a subjective vertical conflict.

1. Introduction

Drivers receive both acceleration and rotational stimulation when negotiating a curve, and it is known that they will tilt their head toward the center of the curve. There are various possible interpretations of the reason for this; for example, it may be to obtain fine visual information about the road geometry (Zikovitz and Harris, 1999). It has also been found that the head movement of passengers is opposite to that of the driver, that is, they tilt their heads in the direction of the centrifugal force (Zikovitz and Harris, 1999). Because passengers tend to be more susceptible to carsickness than are drivers, we assumed that the driver's head movement is related to a decreased likelihood of carsickness and a corresponding increase in comfort. Thus, it is expected that an analysis of drivers' active head-tilt motions can lead to the design of a vehicle with a more comfortable motion.

There are many theories about the possible mechanisms of motion sickness (Golding and Gresty, 2005; Shupak and Gordon, 2006). For instance, it has been postulated that increased instability of postural control increases motion sickness (Riccio and Stoffregen, 1991). The eye-movement hypothesis states that motion sickness can be understood as the result of eye movements controlled by the vestibular system (Ebenholtz et al., 1994). Sensory conflict theory (or neural mismatch theory) is well known, and it postulates that conflicting sensory information leads to motion sickness (Reason, 1978; Shupak and Gordon, 2006); this conflict is due to a mismatch between what the individual senses and what was expected. This conflict is thought to be equivalent to the difference between information obtained by the sensory system and the estimated value composed by the efference copy (Reason, 1978). Oman (1982) proposed a mathematical model for sensory conflict theory, in which motion sickness is estimated by connecting the conflict to the averaging process. The structure of this model, in which the conflict is thought to be at the sensory level, agrees with the evidence of neurobiology (Cullen, 2004). Bles et al. (1998) proposed subjective vertical conflict (SVC) theory, in which the error between the sensed and the estimated vertical direction is considered as the source of the sensory conflict. Bos and Bles (1998) proposed a mathematical model of SVC theory for one-dimensional passive motion. Kamiji et al. (2007) expanded this method to six-degree-of-freedom (6-DOF) motion in three-dimensional (3D) space; they incorporated the semicircular canal and the canal-otolith interaction to allow us to deal with head tilt. They also incorporated a measure of the accuracy of the body motion sensations, including the effect of the efference copy, to deal with the difference between drivers and passengers in the degree of control of the vehicle. Given this mathematical model, it is possible to estimate the motion sickness incidence (MSI), which is defined as the percentage of individuals who vomited due to a given motion stimulus (McCauley et al., 1976). The model uses information about the acceleration and angular velocity of the head to estimate the MSI for a given head motion.

Rolnick and Lubow (1991) discussed the potential factors contributing to the difference in motion sickness between drivers and passengers, namely, controllability, perceived control, activity, visual information, and predictability. Drivers can anticipate their motion in the near future by observing the road geometry and by their active participation in controlling the vehicle. This results in a difference between drivers and passengers in the conflict between sensed information and its estimation from the efference copy. This is similar to the hypothesis of Reason (1978), which is that the difference in the susceptibility to motion sickness occurs because the driver possesses an efference copy related to vehicle control. Visual information also affects anticipation of the motion of the vehicle, and it is known that the wider visual field available to a passenger riding in the front seat also decreases motion sickness (Griffin and Newman, 2004). Also, as mentioned above, drivers tilt their head toward the center when negotiating a curve, whereas passengers tilt their head in the opposite direction. Wada et al. (2012, 2015) demonstrated that when passengers tilt their head toward the centripetal acceleration, imitating the tilt of the drivers' head, this significantly decreases motion sickness. However, there have been no studies of the effect of the driver's head tilt from the viewpoint of sensory conflict theory.

Therefore, the purpose of this study is to analyze the effect of head tilting by drivers and passengers. To do this, we will use the mathematical model proposed by Kamiji et al. (2007), which is based on SVC theory. First, we introduce a mathematical model for head movement in 3D space. Then, the effect of head movement on motion sickness is investigated by applying the mathematical model to both simulated head movements and head movements measured in experiments with real cars.

2. Mathematical Model of MSI Based on Subjective Vertical Conflict Theory

2.1 Subjective vertical conflict: Theory and a mathematical model

The sensory rearrangement theory of motion sickness is well known (Reason and Brand, 1975; Reason, 1978). This theory postulates that situations that provoke motion sickness are characterized by conditions of sensory rearrangement in which there is a conflict between the motion signal transmitted by the visual, vestibular, and somatosensory systems and that which is expected from previous experiences for adapting to new environments. Bles et al. (1998) proposed SVC theory, in which the error between the sensed and estimated vertical or gravitational directions is the source of the conflict. That is, the conflict is defined as the discrepancy between the vertical direction sensed by the vestibular system and the estimate of the vertical direction by the internal model (Merfeld, et al., 1999) that is thought to be built in the central nervous system to resolve ambiguities in the neural signals related to sensory processing and motor control. Bos and Bles (1998) proposed a mathematical model of SVC theory for 1-DOF passive motion. We evaluated the validity of the model by using it to estimate the MSIs for various frequencies and accelerations and then comparing the results of McCauley et al. (1976). Note that the model considers only one-dimensional head motion in the vertical direction without rotation. In addition, the model does not include the effect of the efference copy, which was included in Oman (1982), because the model focuses on passive motions.

2.2 Modeling of MSI for 6-DOF head motion

The 6-DOF-SVC model was proposed for motion sickness due to motion in 3D space (Kamiji et al., 2007), as shown in Fig. 1. This model is an extended version of that of Bos and Bles (1998), and it was created by adding the semicircular canal and canal-otolith interaction to the model of Bos and Bles (1998). This allowed the model to include rotation of the head and a block that can be used to determine the accuracy of the sensation of motion, including the somatic sensation and/or the effect of the efference copy.

The inputs to the model are the gravitoinertial acceleration (GIA) $f$, which is defined as

$$f = a + g, \qquad (1)$$

where $g$ is the acceleration due to gravity and $a$ is the inertia; note that the resultant acceleration of gravity ($g$) and inertia ($a$) work on the otolith of the ear at the same time, according to Einstein's equivalence principle. The vector $f$ is input to the otolith, and this is represented by the block marked OTO in Fig. 1. The transfer function of the OTO is given by a unit matrix. The vector $\omega$ is input to the semicircular canals, which are represented by the block marked SCC in Fig. 1. The transfer function of the semicircular canal that calculates the sensed angular velocity of the head $\omega_s := [\omega_x, \omega_y, \omega_y]^T$ from the angular velocity $\omega$, is given as follows (Merfeld, 1995):

$$\omega_s^j = \frac{\tau_d \tau_a s^2}{(\tau_d s + 1)(\tau_a s + 1)} \omega^j \quad (i = x, y, z), \qquad (2)$$

where $\tau_a$ and $\tau_d$ are time constants.

The sensed vertical direction in the head-fixed frame $v_s$ is estimated from the canal-otolith interaction (Bos and Bles, 2002), as eq.(3):

$$\frac{dv_s}{dt} = \frac{1}{\tau}(f - v_s) - \omega_s \times v_s, \qquad (3)$$

where the time constant is $\tau = 5$ s (Bos and Bles, 2002) and is represented by LP in Fig. 1. In the figure, the lower part of the block diagram shows an internal model of the vestibular system that is thought to exist in the central nervous system. Blocks $\overline{OTO}$ and $\overline{SCC}$ denote the internal models of OTO and SCC, respectively. The transfer function of $\overline{OTO}$ is given by a unit matrix. The transfer function of $\overline{SCC}$ is given as (Merfeld, 1995)

$$\hat{\omega}_s^i = \frac{\tau_d s}{\tau_d s + 1}\tilde{\omega}_i \quad (i = x, y, z), \qquad (4)$$

where $\tau_d$ denotes the time constant, which is the same as that used in eq. (2). Gains $K_a$ and $K_\omega$ represent the estimation errors of the acceleration and the angular velocity, respectively, and these can be understood as being obtained from somatic sensations and/or the effect of an efference copy (Latash, 1998). In the model, we let $K_a = 0.1$ and $K_w = 0.8$, as shown in Table 1. The internal model of LP, which is illustrated as $\overline{LP}$, is assumed to be identical to LP. The outputs of the internal model are $\hat{a}_s$, $\hat{v}_s$, and $\hat{\omega}_s$. The vectors $\Delta a$, $\Delta v$, and $\Delta \omega$ denote the error between the sensory information obtained by the vestibular system, for example, $a_s$, and the information estimated by the internal model, for example, $\hat{a}_s$. These discrepancies are decreased by integration with gains $K_{\omega c}$, $K_{vc}$, and $K_{ac}$. Finally, the MSI is calculated by using the error between the sensed and estimated vertical direction $\Delta \mathbf{v}$, and the Hill function $(\|\Delta \mathbf{v}\|/b)^2 / \{1+(\|\Delta \mathbf{v}\|/b)^2\}$, which models the nonlinear relationship between the MSI and the magnitude of the vertical conflict and the second-order lag with a large time constant $P/(\tau_I s+1)^2$, as depicted in Fig. 1 (Bos and Bles, 1998). The validity of the model was examined by comparing it with experimental results reported in the literature (Donohew and Griffin, 2004); the moving base was vibrated for 2 hours at frequencies from 0.0315 Hz to 0.8 Hz and with a root-mean-square (RMS) acceleration of 0.1g. The calculated MSI was similar to the measured mild nausea reported in the experiments. The results of the proposed method with head rotation fit the experimental results better under the low-frequency conditions than did the results using the model of Bos and Bles (1998), which did not include head rotation. Please refer to Kamiji et al. (2007) and Wada et al. (2010) for more details of the model. All parameters used in the model are presented in Table 1.

Table 1 6-DOF-SVC model parameters

| $K_a$ | $K_w$ | $K_{wc}$ | $K_{vc}$ | $K_{ac}$ |
|---|---|---|---|---|
| 0.1 | 0.8 | 5.0 | 5.0 | 1.0 |
| $\tau_d$ [s] | $\tau_a$ [s] | $b$ [m/s²] | $\tau_I$ [min] | $P$ [%] |
| 7.0 | 190.0 | 0.5 | 12.0 | 85.0 |

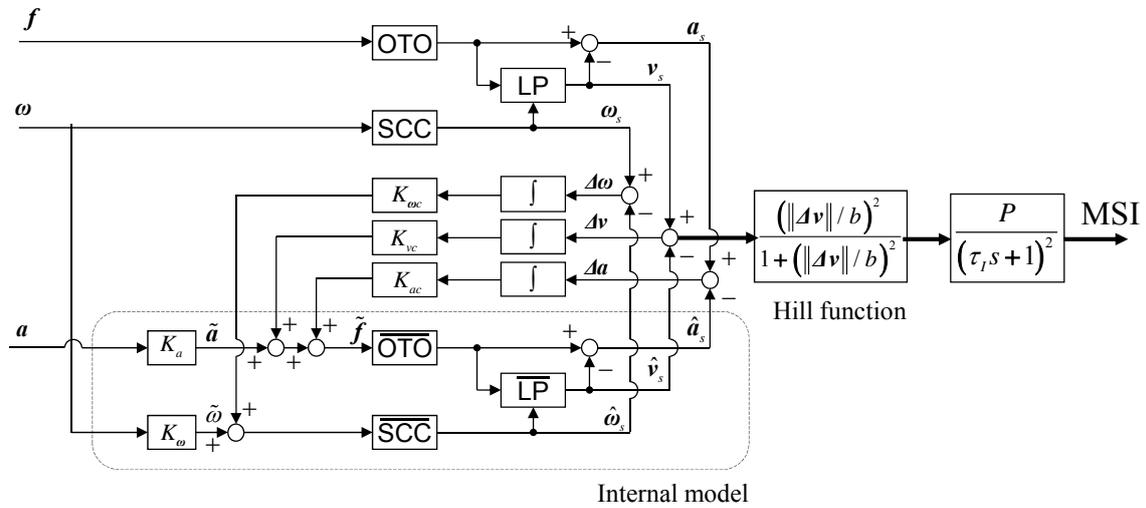

Fig. 1 6-DOF-SVC model of motion sickness incidence, revised from

Kamiji et al. (2007)

3. Predicted MSI with Lateral Acceleration by Simulated Vehicle Motion

The MSI when driving in a slalom course was calculated by using the proposed model with different head movements. The CarSim software package (Mechanical Simulation Corp., Ann Arbor, MI) was used to calculate the simulated vehicle motion; the preview driver model in the CarSim with 0.5 s preview time and 0 s dead time drove a sedan of a 2.37 m wheelbase with a generic setting in slalom track with pylons placed at intervals of 15 m, and speed 30 km/h.

3.1 Simulation with representative head motions

We considered four conditions for the head movements of the driver (see Fig. 2):

1) Compliant condition: the head movements coincide with the vehicle motion; that is, the head is fixed to the vehicle. This imitates the head movements that are sometimes made by passengers.

2) Vertical condition: the head is always vertical, regardless of the motion of the vehicle.

3) Resistant condition: the movement of the head is twice the size and in the opposite direction to the vehicle roll. This approximates experimental observations, as described in the following section.

4) GIA condition: the head is always directed along the GIA vector, which is the sum of the gravitational and lateral accelerations.

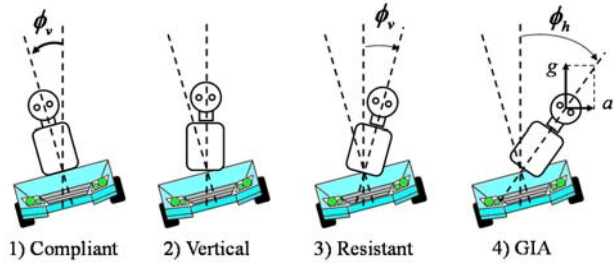

Fig. 2 Head movement conditions

Let the *x*, *y*, and *z* axes be defined as the forward, left, and vertical directions, respectively. In addition, the roll, pitch, and yaw angles of the vehicle with respect to the inertial coordinate system are represented as $[\phi_v \; \theta_v \; \psi_v]^\mathrm{T}$. Then, the angular velocity vector of the head $\omega_h$ in each condition is as follows:

1) Compliant condition: $\omega_h = [\omega_h^x \; \omega_h^y \; \omega_h^z]^\mathrm{T} = [\omega_v^x \; 0 \; \omega_v^z]^\mathrm{T}$,

2) Vertical condition: $\omega_h = [\omega_h^x \; \omega_h^y \; \omega_h^z]^\mathrm{T} = [0 \; 0 \; \omega_v^z]^\mathrm{T}$,

3) Resistant condition: $\omega_h = [\omega_h^x \; \omega_h^y \; \omega_h^z]^\mathrm{T} = [-\omega_v^x \; 0 \; \omega_v^z]^\mathrm{T}$,

4) GIA condition: $\omega_h = \begin{bmatrix} \omega_h^x & \omega_h^y & \omega_h^z \end{bmatrix}^T = T \begin{bmatrix} -\dot{\phi}_h & 0 & \dot{\psi}_v \end{bmatrix}^T$,

where the matrix $T$ is the rotation matrix from the time derivative of the roll, pitch, and yaw angles to the angular velocity. The roll angle of the head $\phi_h$ in the GIA condition is determined such that the $z$ axis of the head coincides with the GIA vector, that is, $\phi_h = \pi/2 - \tan^{-1}(g/a)$, where $g$ and $a$ denote the magnitude of gravitational and lateral acceleration, respectively, of the vehicle shown in the inertial coordinate system. The GIA condition was considered because the direction of the GIA seen in the head-fixed frame does not change, but its magnitude does, and thus it was expected that there would be a smaller discrepancy between it and that estimated by the internal model.

Fig. 3 shows the resultant vehicle motions calculated in the simulation. Note that the lateral acceleration was that of the center of gravity of the vehicle seen from the vehicle-fixed frame. Fig. 4 shows the head movement for each condition associated with the simulated vehicle motion.

The predicted MSI at each moment was calculated using the proposed 6-DOF-SVC model; the input was the time series of the angular velocity of the head $\omega_h$ for each of the four simulation conditions; as discussed above, the inertial acceleration $a$ and GIA $f = g + a$ were determined by the results of the vehicle motion simulation, as shown in Fig. 3. Note that these signals were based on the head-fixed coordinate frame.

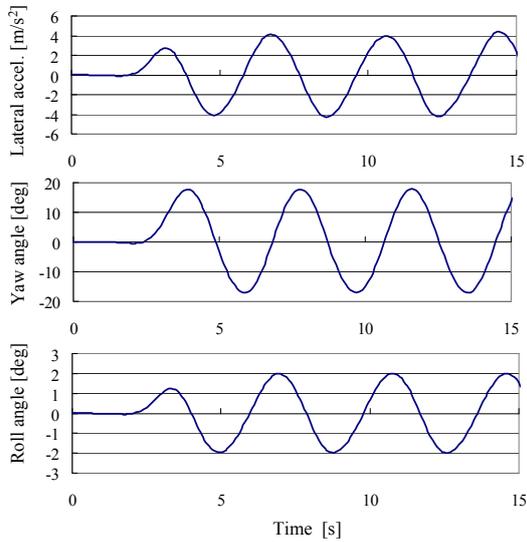

Fig. 3 Resultant vehicle motion in the simulations

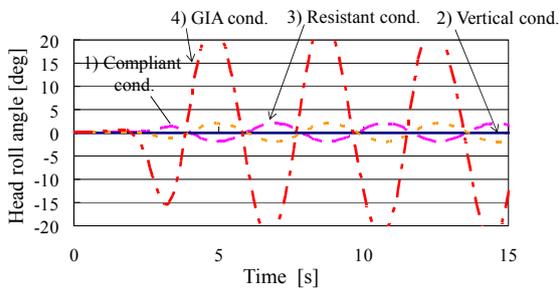

Fig. 4 Resultant head movements in the simulations

Fig. 5 shows the simulation results of the predicted MSI using the 6-DOF-SVC model for 30 min of driving, then stopping for 20 min. The MSI gradually increases over time, and it peaks following the termination of the motion; this is due to the large time constant in the accumulation phase. The MSI was smallest in the GIA condition, even though the maximum head roll angle, which is almost 20 deg, seems larger than that observed in actual drivers (Fig. 8). The MSI in the resistant condition, which is similar to the head movement of actual drivers, was smaller than that of either the compliant or the vertical condition.

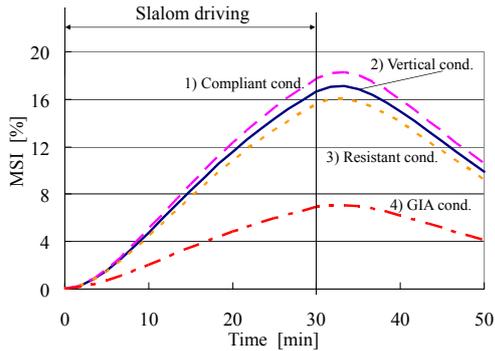

Fig. 5 Time series of simulated MSI under slalom driving conditions

3.2 Comparison of the peak MSI for various head roll angles

Suppose that the head moves as in eq. (4): the amplitude of the head roll angle $\phi_h$ changes from -30 to 50 deg. In this simulation, the vehicle motion is the same as that used in Section 3.1. The compliant, vertical, and resistant conditions defined in the previous section correspond to $\phi_h$ = -2, 0, and +2 deg, respectively; in the GIA condition, $\phi_h$ = 20 deg.

Fig. 6 shows the relationship between the peak MSI and the maximum head roll angle. The minimum MSI values are at approximately 30 deg, even though such a head roll angle seems larger than that observed in the real world. From this simulation, it was found that actively rolling the head in the direction opposite to the lateral acceleration and vehicle roll reduces the MSI.

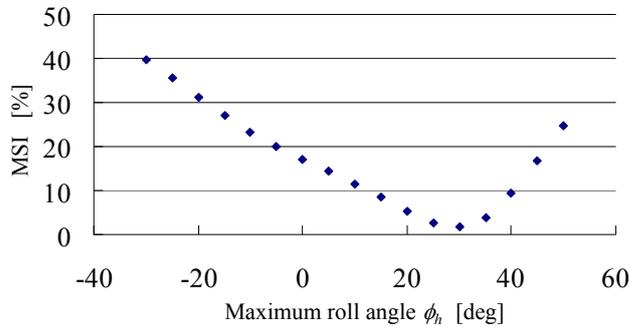

Fig. 6 Simulated peak MSI under slalom driving conditions

## 4. MSI Predicted from Measured Head Movements of Drivers and Passengers

### 4.1 Experiments with a real car

#### 4.1.1 Apparatus

An experiment with human subjects in a real car was conducted to measure the driver and passenger head movements, and then the estimated MSI was calculated by using the 6-DOF-SVC model.

A small passenger car with a 2.46 m wheelbase and a 1300 cc engine was used for the driving experiments. A motion sensor (MTi-G, Xsense Technologies, Enschede, Netherlands) was attached to a flat place close to the shift lever of the automatic transmission in order to measure the 3-DOF acceleration and the 3-DOF orientation of the vehicle (Fig. 7). A gyro-type orientation sensor (Inertia Cube3, InterSense, Billerica, MA) and a wireless accelerometer (WAA-001, Wireless Technology, Ventura, CA) were attached to the cap worn by the participants in order to measure the 3-DOF orientation and the 3-DOF acceleration, respectively, of the head in the head-fixed coordinate frame (Fig. 7).

These sensors were connected to a laptop PC in the rear seat of the vehicle in order to synchronize the sensor data. A straight asphalt road was used as the test track, and five pylons were located at regular distances of either 15 m or 20 m.

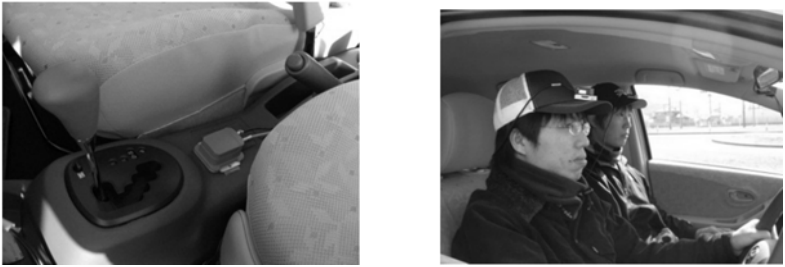

Fig. 7 Sensors attached to the vehicle and the participants' heads

4.1.2 Design

Six males aged 22 to 24 yr who had driver's licenses and who gave informed consent participated in the experiments as both drivers and passengers. The driver/passenger condition was treated as a within-subject factor. Each driver was asked to drive the pylon slalom on the straight road at a constant velocity. There were two conditions in the slalom driving, as follows:

Mild condition: Driving at 30 km/h with a 20 m gap between pylons,

Hard condition: Driving at 40 km/h with a 15 m gap between pylons.

A second participant sat in the front passenger seat. The order of the slalom conditions was randomized among the participants. Note that the nominal frequencies of lateral acceleration during slalom driving were 0.21 Hz (the most provocative frequency) in the mild condition and 0.37 Hz in the hard condition.

4.1.3 Procedure

In each experiment, the driver was seated in the car in the normal driving position, and the passenger was seated in the front passenger seat in the normal riding position; both wore safety belts. The experimental course was a straight road with five pylons, and the two subjects were exposed to acceleration as the car was driven through the course. The drivers were asked to drive at a predetermined constant velocity. Following some practice runs, each driver then drove the test track three times per each slalom driving condition. The RMS lateral acceleration of the vehicle was 1.83 m/s$^2$ (SD 0.16) in the mild condition and 3.45 m/s$^2$ (SD 0.64) in the hard condition.

4.2 Results

4.2.1 Vehicle motion and head roll

Figs. 8 and 9 show examples of the lateral acceleration and the roll angle of the vehicle and those of the driver's and passenger's heads in the mild and hard conditions, respectively. It was found that the driver's head moved in the direction opposite to the roll of the vehicle. The passenger's head moved passively in either the direction opposite direction to that of the driver or in the same direction but with a large time lag; thus, this motion was not in good synchronization with the vehicle motion. The maximum head roll angles of the drivers and the passengers in the hard condition were larger than those in the mild condition.

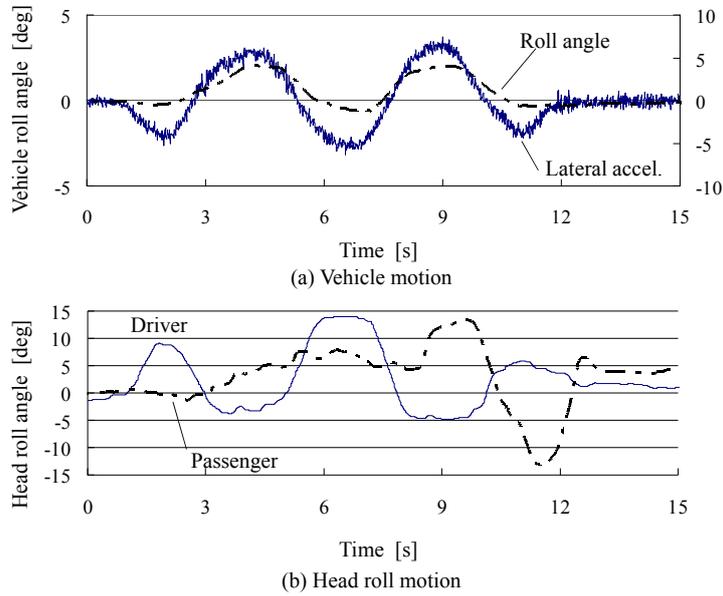

Fig. 8 Vehicle and head motions in the experiments (mild condition)

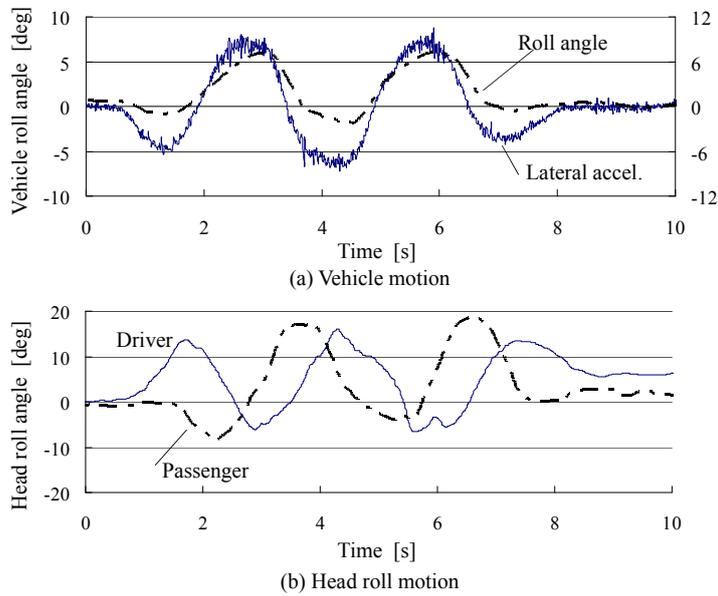

Fig. 9 Vehicle and head motions in the experiments (hard condition)

Fig. 10 shows the RMS head roll angles. The Wilcoxon signed-rank test for each slalom condition revealed that there were no significant differences between the driver and passenger in either the mild condition ($z = 0.524$, $n = 6$, ties = 0, $p = 0.600$, two-tailed) or the hard condition ($z = 0.314$, $n = 6$, ties = 0, $p = 0.753$, two-tailed). In addition, no significant differences were found in the head roll angle between the mild and hard conditions ($z = 0.628$, $n = 12$, ties = 0, $p = 0.530$, two-tailed).

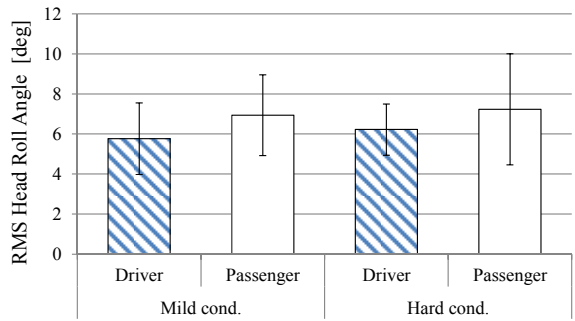

Fig. 10 RMS head roll angles during slalom driving in the experiments

The correlation coefficients between the vehicle lateral acceleration and the head roll angle during the slalom driving were analyzed. Fig. 11 shows the mean correlation coefficients for all participants. The error bars indicate the standard deviation. A positive correlation coefficient indicates that the head roll motion synchronized well with the vehicle lateral acceleration in the centrifugal direction. A high negative correlation was found in the driver results, while a lower positive correlation was found in the passenger results. The Wilcoxon signed-rank test revealed a significant difference between the drivers and the passengers in the mild condition ($z = 2.20$, $n = 6$, ties = 0, $p = 0.028$, two-tailed) and the hard condition ($z = 2.20$, $n = 6$, ties = 0, $p = 0.028$, two-tailed). A measure of the effect size *r* was 0.898 for both conditions. In both conditions, the head roll angle of the drivers was greater in the centripetal direction, whereas the head movements of the passengers were occurred in both the centripetal and centrifugal directions (Fig. 12).

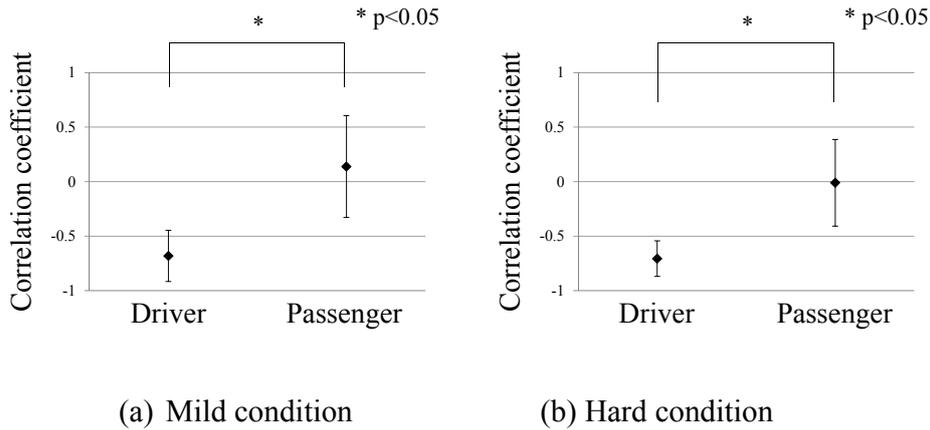

(a) Mild condition   (b) Hard condition

Fig. 11 Correlation coefficients between the vehicle lateral acceleration and the head roll angle

4.2.2 Prediction of the MSI based on measured head roll motions

The predicted MSI was calculated from the head motions measured in the experiments; the angular velocity of the head was used as input to the 6-DOF-SVC model. Due to the large amount of noise in the measurements of the acceleration of the head of the driver and passenger, the acceleration of the vehicle was used in all conditions. The MSI was then calculated using the experimental results obtained during slalom driving.

Fig. 12 shows the results of the predicted MSI for each condition. The Wilcoxon signed-rank test revealed the significance of the difference between the drivers and the passengers in the mild condition ($z = 1.99$, $n = 6$, ties $= 0$, $p = 0.046$, two-tailed); the effect size *r* was found to be 0.812. The predicted MSI was calculated from the head movements; that of the drivers was smaller than that calculated for the passengers. In contrast, in the hard condition, no significant difference was found between the MSI of the driver and that of the passenger ($z = 1.36$, $n = 6$, ties $= 0$, $p = 0.173$, two-tailed).

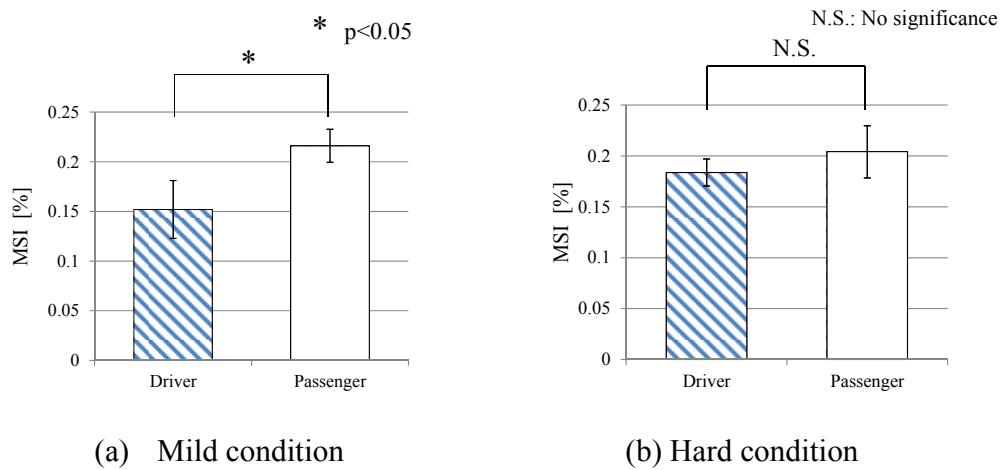

(a) Mild condition  (b) Hard condition

Fig. 12 Predicted MSI calculated from the experimental results

4.2.3 Changes due to gains in the MSI predicted by 6-DOF-SVC

As mentioned above, the gains $K_a$ and $K_w$ represent the accuracy of the sensation of self-motion, including somatic sensation and/or the effect of the efference copy, which is strongly related to the controllability of motion. Thus, the predicted MSI of the driver's head movement was calculated with other gain settings ($K_a = 0.5$ and $K_w = 0.9$) in order to investigate the effect of the efference copy on the MSI. The mean MSI (and SD) in the mild and hard conditions were 0.0909% (SD 0.038) and 0.123% (SD 0.011), respectively. The Wilcoxon signed-rank test revealed that the predicted MSI decreased when the gains increased, for both the mild condition ($z = -2.201$, $n = 6$, ties = 0, $p = 0.028$, two-tailed) and the hard condition ($z = -2.2201$, $n = 6$, ties = 0, $p = 0.028$, two-tailed). The effect size $r$ in each case was 0.899.

5. Discussion

In the experiment with a real car, the drivers tilted their head toward the centripetal direction, while the direction in which the passengers tilted their head had a large variance and a low mean correlation with the lateral acceleration of the vehicle. This implies that sometimes they tilted their head in a direction similar to that of the driver, and in some cases, they tilted their head in the opposite direction. Our findings for the driver are consistent with the qualitative observations of the head movements of bus drivers (Fukuda, 1976) and with other experiments in actual cars (Zikovitz and Harris, 1999). Our findings for the passenger are not consistent with those of Fukuda (1976) and Zikovitz and Harris (1999), since they found that passengers always tilted their head in the direction opposite to that of the driver. The reason for this inconsistency is thought to come from differences in the experimental conditions. In the present study, we used slalom driving, in which the lateral acceleration changes, and thus the passenger is forced to continuously change their control strategy; however, in the previous studies, a normal curved road was used (Fukuda, 1976; Zikovitz and Harris, 1999).

We used the 6-DOF-SVC model to predict the MSI, using the experimental results as input. In the mild condition, the MSIs predicted for the drivers were significantly smaller than those predicted for the passengers. We also found that for the motions generated by the simulated slalom driving, if we assumed that the head was tilted in the centripetal direction, the MSI was reduced. This agrees with simulation results in the literature (Wada et al., 2010), which also predicted the MSI using the 6-DOF-SVC model; it was found that when riding on a base that moved sinusoidally in the lateral direction, the predicted MSI is reduced when the head is tilted against the inertial acceleration, and thus it is aligned more closely with the GIA. The MSI was originally defined as the percentage of individuals who vomited in response to a given stimulus (McCauley, et al., 1976). An increase in the MSI through a change in the stimulus means that some of the individuals who did not vomit when presented with the previous stimulus, vomited in response to the new stimulus. In many cases, individuals vomited after experiencing initial symptoms such as pale skin, cold sweat and nausea, although there were a few exceptions. Therefore, assuming that the distribution of the motion sickness susceptibility of the subjects used in the MSI calculation was similar to that of the human population at large, changes in MSI can be regarded as the changes in severity of motion sickness experienced by average individuals. In addition, Wada et al. (2010) demonstrated that the MSI predicted by our mathematical model shows a distribution similar to the percentage of mild nausea observed by Donohew & Griffin (2004). Based on these considerations, we regard the MSI predicted by our model as an index of the severity of motion sickness. With this interpretation, these findings also agree with the results of Golding et al. (2003), in which the severity of motion sickness decreased when the head was actively aligned with the GIA in a longitudinal linear acceleration environment. This result suggests that for passengers, the severity of motion sickness can be reduced by tilting their head toward the centripetal direction, as is done by drivers. The results of experimental studies with real cars by Wada et al. (2012) and Wada and Yoshida (2015) are in agreement with this idea; those studies also revealed that the severity of motion sickness was reduced significantly when the passenger imitated the driver's head movements during slalom driving. The contribution of the present paper is that it demonstrates the effect of head tilt on reducing the severity of motion sickness in a real automotive environment under lateral acceleration from the viewpoint of SVC theory. We note that the estimated MSI for the driver's head movement was smaller than that for the passenger's head movement, even when the gains, $K_a$ and $K_w$, were the same; these represent the accuracy of perception of body movement or spatial orientation, including the effect of the efference copy. We also note that this was not found to be significant in the hard condition. It can be understood that this is either because it is difficult for drivers to tilt their head appropriately during the hard condition, or because the passengers need to control their posture against the hard lateral acceleration, and this leads to a head tilt in a direction similar to that of the drivers. Furthermore, for the drivers, an increase in the gains led to a smaller predicted MSI, even with the same head movement. This implies that the head movement itself has the effect of reducing motion sickness, and the MSI can be further decreased by increasing the accuracy of the perception of the body movement, including the effect of the efference copy. These results are strongly related to controllability and predictability, which were noted by Rolnick and Lubow (1991) as two of the important factors related to motion sickness, and these factors are greater for the driver than for the passengers. In addition, for active movement, adaptation to the motion is promoted by comparing the efference copy of the motor command with the reafference from the effector, according to the reafference principle (von Holst, 1954). Assuming that these gains represent the accuracy of the perception of one's motion and is related to controllability and predictability or to the efference copy, this suggests that the 6DOF-SVC model can determine the decrease in the MSI when an individual actively participates in the control of the vehicle. This is in addition to the effect of the head movement. Relationship between the predicted MSI and the severity of motion sickness should be investigated for further understanding of the difference of motion sickness among conditions.

It should also be noted that the 6-DOF-SVC model does not include visual or somatosensory information, because it is limited to the vestibular sensory system. However, the experimental results of Wada and Yoshida (2015) showed that the effect of reducing motion sickness by tilting the head in the centripetal direction is valid whether the eyes are open or closed. This implies that the 6-DOF-SVC model could predict the MSI for various head-tilt strategies, regardless of whether the eyes are open or closed. We do not have a rationale for assigning a quantitative value to the change in the gains that represent the accuracy of perception of body movements. This is a limitation of the present paper.

6. Conclusions

The effect of the driver's head movement on the incidence of motion sickness was investigated using the 6-DOF-SVC model in simulations and in actual cars with actual drivers and passengers. The results revealed that the estimated MSI due to the driver's head movements was smaller than that of the passenger's movements. It was shown that by changing the gains in the 6-DOF-SVC model to increase the accuracy of the perception of body movements, which is related to somatic sensation and the efference copy, a lower MSI is predicted. From this result, it was suggested that the driver's head tilt toward the centripetal direction has the effect of reducing motion sickness, and the predictability and controllability experienced by the driver further decease motion sickness. These results give us a new interpretation of the driver's head movements; that is, we can use SVC theory to explain why particular head movements reduce motion sickness. This suggests that the severity of the MSI of the passengers can be reduced by matching the drivers and tilting their head toward the centripetal direction.

   Visual information needs to be introduced to the 6-DOF-SVC model, because the current version is limited to the vestibular sensory system. A conceptual model of SVC theory with visual input was postulated by Bos et al. (2008). A mathematical model including visual input can be utilized to investigate the effect of the driver's behavior in more detail. These efforts will increase basic knowledge and can thus result in better designed vehicles with more comfortable motion. The relationship between the predicted MSI and the severity of motion sickness should be investigated for further understanding of the differences in the motion sickness experienced in a variety of conditions.

Acknowledgements

The authors would like to thank Mr. Hiroyuki Konno for his help in data collection. The authors also would also like to thank the anonymous reviewers for their valuable comments and suggestions for improving the quality of this paper. This research did not receive any specific grant from funding agencies in the public, commercial, or not-for-profit sectors.